\documentclass[epj]{webofc}
\usepackage[varg]{txfonts}   
\woctitle{International Nuclear Physics Conference 2013}
\begin{document}
\title{PROBING THE STATISTICAL DECAY AND $\alpha$-CLUSTERING 
\\EFFECTS IN $^{12}$C+$^{12}$C AND $^{14}$N+$^{10}$B REACTIONS.}
%
%

\author{L. Morelli\inst{1}\fnsep\thanks{\email{luca.morelli@bo.infn.it}} \and
        G. Baiocco\inst{2}\and
        M. D’Agostino\inst{1}\and
M. Bruno\inst{1}\and
F. Gulminelli\inst{3}\and
M. Cinausero\inst{4}\and
M. Degerlier\inst{5}\and
D. Fabris\inst{6}\and
F. Gramegna\inst{4}\and
T. Marchi\inst{4}\and
S. Barlini\inst{7}\and
M. Bini\inst{7}\and
G. Casini\inst{7}\and
N. Gelli\inst{7}\and
A. Lopez\inst{7}\and
G. Pasquali\inst{7}\and
S. Piantelli\inst{7}\and
S. Valdrè\inst{7}.
}

\institute{Dipartimento di Fisica e Astronomia dell’Università and INFN, Sezione di Bologna, Bologna, Italy.
\and
           Dipartimento di Fisica dell’Università and INFN, sezione di Pavia, Pavia, Italy. 
\and
           CNRS, UMR6534, LPC, Caen, France and ENSICAEN, UMR6534, LPC, Caen, France.
\and
           INFN, Laboratori Nazionali di Legnaro, Legnaro (Padova), Italy.
\and
           University of Nevsehir, Science and Art Faculty, Physics Department, Nevsehir, Turkey.
\and 
           INFN, Sezione di Padova,, Padova, Italy.
\and
           Dipartimento di Fisica dell’Università and INFN, Sezione di Firenze, Firenze, Italy.
          }

\abstract{%
An experimental campaign has been undertaken at Laboratori Nazionali di Legnaro (LNL INFN), Italy, in order to progress in our understanding of the statistical properties of light nuclei at excitation energies above particle emission threshold, by measuring exclusive data from fusion-evaporation reactions. 
On the experimental side, a first reaction: $^{12}$C+$^{12}$C at 95 MeV beam energy has been measured, using the GARFIELD + Ring Counter (RCo) apparatuses. Fusion-evaporation events have been exclusively selected out of the entire data set. The comparison to a dedicated Hauser-Feshbach calculation allows us to give constraints on the nuclear level density at high excitation energy for light systems ranging from C up to Mg. Out-of-equilibrium $\alpha$ emission has been evidenced and attributed both to an entrance channel effect (favoured by the cluster nature of reaction partners), and, in more dissipative events, to the persistence of cluster correlations well above the $^{24}$Mg threshold for 6 $\alpha$'s decay. In order to study the same $^{24}$Mg compound nucleus at similar excitation energy with respect to this first reaction a new measurement, $^{14}$N + $^{10}$B at 5.7 A.MeV, was performed at LNL laboratories with the same experimental setup. The comparison between the two systems would allow us to further constrain the level density of light nuclei in the mass-excitation energy range of interest. In this perspective, deviations from a statistical behaviour can be used as a tool to get information on nuclear clustering, both in the ground-state for projectile and target and in the hot source formed in the collision.

}
\maketitle
\section{Introduction}
\label{intro}
The NUCL-EX collaboration has recently proposed an experimental campaign of exclusive measurements of fusion-evaporation reactions with light nuclei as interacting partners. The aim is to progress in the understanding of statistical properties of light nuclei decay at excitation energies above particle emission thresholds. 
These properties notably include the excitation energy dependence of the nucleon effective mass, symmetry energy and pairing correlations. In particular, the determination of the nuclear level density in the A$\sim$20 region, the understanding of the statistical behaviour of light nuclei with excitation energies $\sim$3 A.MeV, and the measurement of observables linked to the presence of cluster structures of nuclear excited levels are the main physics goals of this work.
The choice of investigating light systems is due to the fact that the low fragments multiplicity produced in these collisions increases the probability of achieving a quasi-complete reconstruction of the event.  
\\Moreover some excited states of different nuclei in this mass region are known to present pronounced cluster structures; these correlations may persist in the ground state along some selected isotopic chains \cite{freer}; according to the Ikeda diagrams \cite{ikeda} alpha-clustered excited states are massively expected at high excitation energies close to the multi-alpha decay threshold in all even-even N = Z nuclei. For such light systems, signatures of cluster structure in the reactions are therefore expected to be more evident even at high excitation energy.
\\In the framework of this campaign, the $^{12}$C+$^{12}$C and $^{14}$N + $^{10}$B reactions have been measured in order to study the decay of the same $^{24}$Mg compound nucleus, populated at the same excitation energy $\varepsilon^{*}= 2.6$ A.MeV but through different entrance channel.

\section{The experiments}
\label{sec-2}
The experiments were performed at the LNL (Laboratori Nazionali di Legnaro), with the $^{12}$C and $^{14}$N beams provided by the XTU TANDEM accelerator.
The experimental setup is composed by the GARFIELD \cite{garf,newgarf} detector and the Ring-Counter (RCo) annular detector \cite{rco}, now fully equipped with digital electronics \cite{elet}, the combination of the two devices allows a nearly-4$\pi$ coverage of the solid angle, which, combined with a high granularity, permits to measure the charge, the energy and the emission angles of nearly all the charged reaction products, allowing an excellent discrimination of the different reaction mechanisms. They also provide information on the mass of the emitted charged products in a wide range of particle energy and type. \\The GARFIELD detector, covering almost completely the angular range of polar angles from 30$^{\circ}$ to 150$^{\circ}$ is a two detection stage device, made by a microstrip gaseous drift chamber ($\mu$SGC), filled with CF$_{4}$ gas at low pressure (about 50 mbar), and CsI(Tl) scintillation detectors lodged in the same gas volume.
The RCo detector is a forward-angle array of three-stage telescopes realized in a truncated cone shape. The first stage is an ionization chamber (IC), the second a 300 $\mu$m reverse-mounted Si(nTD)-detector, and the last a CsI(Tl) scintillator. It has azimuthal symmetric, with 8 sectors, and covers the polar region from $6^{\circ}\leq \theta \leq 18^{\circ}$, with an angular resolution  $\Delta\theta\approx$0.7 and an energy resolution of 0.3$\%$ (silicon strips) and 2-3$\%$ (CsI (Tl)).
\\Due to the reverse mounting of the nTD Silicon detector it has been possible to identify the charge of stopped fragments via discrimination analysis (PSA) with digital electronics, according to the results of the FAZIA detectors\cite{digital,nicola}. In addition, it has been recently proved the feasibility of isotopic separation via pulse shape analysis of silicon detector for fragments with charge up to Z = 14 \cite{newgarf}.
 
\section{Results and discussion}
\label{sec-3}
\begin{figure}
 \centering
\includegraphics[width=0.8\columnwidth]{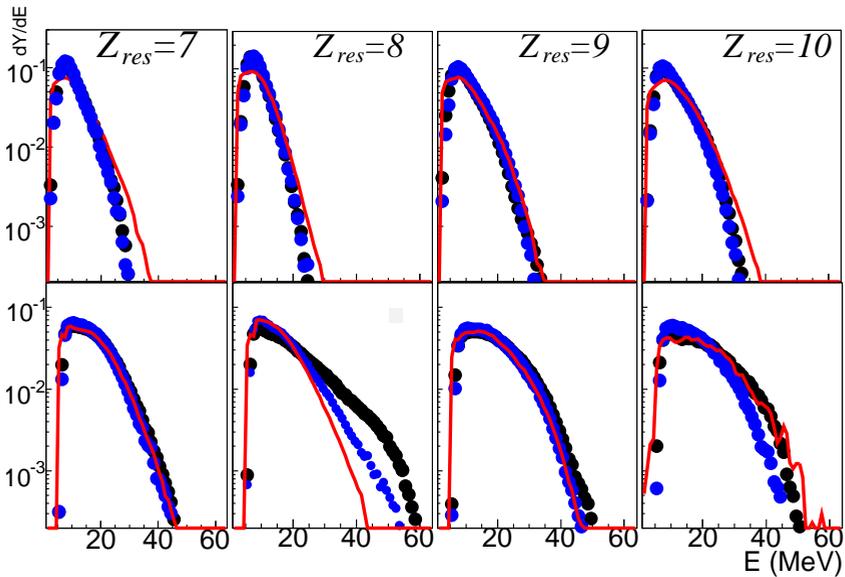}
\caption{(Color online) Proton (upper part) and $\alpha$ (lower part) energy spectra in complete Z$_{det}$=12 events detected in
coincidence with a residue of charge Z$_{res}$, indicated in each figure column. Data (black and blue dots for $^{12}$C+$^{12}$C and $^{14}$N+$^{10}$B respectively) are compared to model calculations (red lines).}
 \label{fig-1}       
 \end{figure}

The fusion-evaporation channel can be selected setting conditions on the total detected charge and on the coincidence between a residue at forward angles (RCo) ($6^{\circ}\leq \theta \leq 18^{\circ}$) and light charged particles (LCP) detected in GARFIELD ($\theta_{lab}\geq30$). Only complete events where the total charge of the entrance channel (Z$_{det}$ = 12) is detected are retained for the analysis.
\\Experimental data are therefore compared to the predictions of a Monte Carlo Hauser-Feshbach code \cite{hf} for the evaporation of the compound nucleus $^{24}$Mg, at $\varepsilon^{*}$ = 2.6 A.MeV, corresponding to a complete fusion source, and filtered through a software replica of the experimental set-up.
With this preliminary selection, the energy spectra of protons and $\alpha$ particles detected at GARFIELD angles are shown in Fig.~\ref{fig-1}, for the $^{14}$N reaction and compared to HF \cite{hf} calculations and data from $^{12}$C+$^{12}$C experiment.
\\For both reactions, a good reproduction of the proton and $\alpha$ energy spectra is achieved in all channels, but in the oxygen case, the energy tails for alpha are not reproduced by the model.

\begin{figure}
 \centering
 \includegraphics[width=0.8\columnwidth]{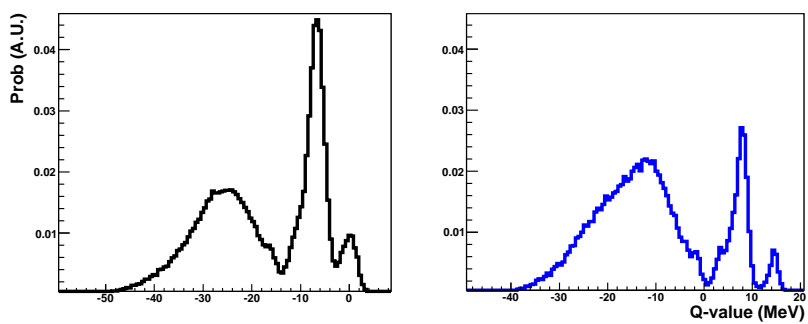}
 \caption{Q-values of the channel $^{A}$O + $\alpha$ + $\alpha$ for $^{12}$C+$^{12}$C (left) and $^{14}$N+$^{10}$B (right). For more details see text.}
 \label{fig-2}       
 \end{figure}

It has been shown in \cite{gio} that the discrepancy observed in Fig.\ref{fig-1} between $^{12}$C+$^{12}$C data and HF calculations for $\alpha$ particles in coincidence with O residues is mostly due to the an extra experimental cross section for nondissipative channels of the type (2$\alpha$, $^{16}$O$^{gs/*}$), favoured by the $\alpha$-cluster structure of reaction partners. \\This seems to be confirmed by the new data set, since, with a different entrance channel, the observed deviation appears largely reduced, even if a residual discrepancy is still present. Thanks to the completeness of the event reconstruction, a Q-value distribution can be built to further investigate the (2$\alpha$, $^{A}$O) channel in both reactions: $Q_{kin} = \sum_{i=1}^2 E_{\alpha_{i}} + E_{O} - E_{beam}$. Fig.~\ref{fig-2} displays the obtained $Q_{kin}$ distributions for $^{12}$C+$^{12}$C (left panel) and $^{14}$N+$^{10}$B (right) reactions.

We can see that the two spectra show a common structure, with two narrow peaks and a broader region extending up to a high energy shortage. 
In the statistical model interpretation, the two peaks correspond to $\alpha$-decay chains, starting from the $^{24}$Mg$^*$ compound nucleus and leaving an $^{16}$O residue either in its ground state or in one of its excited bound states which are not resolved in the experiment.
The Q-values $Q_{kin}=-15.78$ MeV for $^{12}$C and $Q_{kin}=-0.8$ MeV for $^{14}$N reactions correspond to the opening of the 4-body channel $^{15}$O+$n+\alpha+\alpha$. Neutrons are not detected in these experiments, and the broader distribution observed for lower $Q_{kin}$ values is due to events in which neutron(s) emission has taken place, and their kinetic energy has not been collected.
\\Besides the common pattern observed for the Q-value distributions, a difference in the relative population of the less dissipative events is evident between the $^{12}$C+$^{12}$C (left panel) and $^{14}$N+$^{10}$B (right) reactions in Fig.~\ref{fig-2}. 
In particular, a much higher percentage of (2$\alpha$,$^{16}$O) events populates the less dissipative Q-value region in the $^{12}$C experimental sample. \\Concluding, the Q-value analysis of the (2$\alpha$, $^{A}$O) channel for the new data-set confirms the possible contamination of direct ($\alpha$-transfer/pick-up) reactions for the $^{12}$C+$^{12}$C reactions. \\Further investigations through a complete channel by channel analysis are needed, in order to understand if the residual deviation from statistical model predictions appearing in the $\alpha$ particle spectrum in coincidence with oxygen in the $^{14}$N+$^{10}$B dataset may be linked to the persistence of cluster correlations at high excitation energy in the fused system, which is the same $^{24}$Mg$^*$ nucleus for both reactions.

%
%

\end{document}